\documentclass[showpacs,amssymb,twocolumn,floatfix,pre,aps]{revtex4-2}
\usepackage{amssymb,amsfonts,bm}
\usepackage{graphics,graphicx}
\usepackage{fontenc}
\usepackage{t1enc}
\usepackage{amsmath}
\usepackage{float}
\usepackage{setspace}
\usepackage{multirow}
\usepackage{color,xcolor}
\usepackage{graphics,psfrag}
\usepackage{graphicx,psfrag}
\usepackage{enumerate}
\usepackage[none]{hyphenat}

\DeclareMathOperator{\Tr}{Tr}

\begin{document}

\title{Full nonuniversality of the symmetric 16-vertex model on the square 
lattice}

\author{Eva Posp\'{\i}\v{s}ilov\'a}
\author{Roman Kr\v{c}m\'ar}
\author{Andrej Gendiar}
\author{Ladislav \v{S}amaj}
\affiliation{Institute of Physics, Slovak Academy of Sciences, 
D\'ubravsk\'a cesta 9, 84511 Bratislava, Slovakia}

\date{\today} 

\begin{abstract}
We consider the symmetric two-state 16-vertex model on the square lattice
whose vertex weights are invariant under any permutation of adjacent 
edge states.
The vertex-weight parameters are restricted to a critical manifold which is 
self-dual under the gauge transformation. 
The critical properties of the model are studied numerically by using 
the Corner Transfer Matrix Renormalization Group method.
Accuracy of the method is tested on two exactly solvable cases: the
Ising model and a specific version of the Baxter 8-vertex model in a 
zero field that belong to different universality classes.
Numerical results show that the two exactly solvable cases are connected by 
a line of critical points with the polarization as the order parameter. 
There are numerical indications that critical exponents vary continuously 
along this line in such a way that the weak universality hypothesis
is violated. 
\end{abstract}

\maketitle

\section{Introduction} \label{Sec1}
According to the universality hypothesis \cite{Griffiths70}, critical exponents 
of a statistical system at the second-order phase transition do not depend 
on details of the corresponding Hamiltonian.
Equivalently, the critical exponents depend only on the system's 
space dimensionality and the symmetry of microscopic degrees of freedom 
(say, the spins).
The first violation of the universality hypothesis was observed in
the Baxter's exact solution of the two-dimensional (2D) 8-vertex model on 
the square lattice in a zero electric field \cite{Baxter71,Baxterbook,Samaj13}
whose critical exponents are functions of model's parameters.
Suzuki \cite{Suzuki74} argued that the violation of universality in the
8-vertex model is due to an ambiguous identification of the deviation 
from the critical temperature. 
If taking, instead of the usual temperature difference $\vert T_c-T\vert$,
the inverse correlation length $\xi^{-1} \propto \vert T_c-T\vert^{\nu}$ 
(the critical exponent $\nu$ is assumed to be the same for both limits
$T\to T_c^-$ and $T\to T_c^+$) as the natural measure of the distance from 
the critical temperature, the renormalized thermal exponents $\alpha/\nu$,
$\beta/\nu$ and $\gamma/\nu$ become universal, i.e., independent of 
the model's parameters.
The critical exponents defined just at the critical temperature, 
such as $\delta=1+\frac{\gamma}{\beta}$ and $\eta=4/(\delta+1)$, stay constant
when varying 8-vertex model's parameters.
This phenomenon is known as weak universality.
Weak universality was observed in many 2D systems, 
including the Ashkin-Teller model \cite{Ashkin43,Kadanoff77,Zisook80},
absorbing phase transitions \cite{Noh04},
the spin-1 Blume-Capel model \cite{Malakis09},
frustrated spin models \cite{Queiroz11,Jin12}, 
percolation models \cite{Andrade13}, etc.
There are few exceptions from models with continuously varying critical 
exponents which violate weak universality, such as 
micellar solutions \cite{Corti82},
Ising spin glasses \cite{Bernardi95},
itinerant composite magnetic materials \cite{Fuchs14,Khan17}, etc.

To set up terminology, the full violation of universality means
that the critical exponents vary continuously as functions of some model's
parameter(s) in such a way that at least one of the renormalized thermal 
exponents $\alpha/\nu$, $\beta/\nu, \gamma/\nu$ or $\delta,\eta$
is nonconstant.
We do not use the term nonuniversality for models which have several 
regions in their parameter space belonging to different universality classes
because the corresponding order parameters are defined differently.

The partition function of the ``electric'' 8-vertex model on the square 
lattice can be mapped onto the partition function of a ``magnetic'' Ising 
model on the dual (also square) lattice with the nearest-neighbor 
two-spin and four-spin interactions on a square plaquette 
\cite{Wu71,Kadanoff71}.
Baxter's exact solution of the zero-field 8-vertex model 
\cite{Baxter71,Baxterbook} provides all magnetic critical exponents
(exhibiting weak universality), but only one electric critical exponent 
(namely $\beta_{\rm e}$ which describes the temperature singularity of 
the spontaneous polarization).
Recently two of us \cite{Krcmar18} argued that the critical exponents
related to the divergence of the correlation length must coincide in
both the magnetic and the electric models: $\nu_{\rm e}=\nu$.
Having two critical exponents at one's disposal, all remaining 
electric exponents can be derived by using scaling relations.
The obtained analytic formulas for the electric critical exponents are
in perfect agreement with numerical results obtained by the Corner Transfer
Matrix Renormalization Group method \cite{Krcmar18}.
It turns out that the model's variation of the electric critical exponents 
violates weak universality.
Thus, despite the partition functions of the electric and magnetic 
models are equivalent, their critical properties are fundamentally 
different: while the magnetic critical exponents obey weak universality,
the electric ones do not and therefore they are fully nonuniversal.  

The partition function of a vertex model is invariant under
gauge transformation of vertex weights \cite{Wegner73,Gaaff75} 
which is a generalization of the weak-graph expansion \cite{Nagle68}
and the duality transformation.
If a point in the parameter space of vertex weights is mappable onto
itself by a nontrivial gauge transformation, that point belongs 
to the self-dual manifold where all critical points of second-order 
phase transitions lie.

The model under consideration in this paper is the symmetric two-state 
16-vertex model on the square lattice whose vertex weights are isotropic,
i.e., invariant under any permutation of the adjacent edge states.
This model was introduced in Ref. \cite{Wu89} in connection with 
the O(2) gauge transformation which preserves the permutation symmetry of
vertex weights and its self-dual manifolds can be easily found.
In a certain subspace of the vertex weights, the model can be mapped 
onto Ising spins in a field \cite{Samaj91,Samaj92}.
The critical properties of the model were studied numerically
by combining a series expansion on the lattice and the Coherent Anomaly
method \cite{Suzuki86} in Ref. \cite{Kolesik93}.
In spite of modest computer facilities and lack of efficient numerical methods 
at that time (almost 30 years ago), the numerical results indicate 
the full nonuniversality of the model. 
(For a recent survey of the general 16-vertex model with an enlargement
of known mappings, see Ref. \cite{Assis17}.)
 
The aim of this work is to revisit the study of the critical electric 
properties of the symmetric version of the 16-vertex model 
on the square lattice by using the Corner Transfer Matrix Renormalization 
Group (CTMRG) method \cite{Nishino96,Nishino97,Ueda05,Krcmar16,Genzor17}.
The method is based on the density matrix renormalization 
\cite{White92,White93,Schollwock05} and the technique of 
the corner transfer matrices \cite{Baxterbook}.
It has been applied to many 2D lattice models and provides very accurate results
for critical points and exponents.
The present work confirms with a high reliability that the symmetric 16-vertex
on the square lattice is nonuniversal and violates the weak universality
hypothesis.

The paper is organized as follows.
The definition and basic facts about the model, including the gauge
transformation of vertex weights, are given in Sec. \ref{Sec2}.
Two exactly solvable cases are discussed: the Ising model and 
a specific version of the Baxter eight-vertex model in zero field.
The CTMRG method is reviewed briefly in Sec. \ref{Sec3}.
Numerical results for the critical temperatures and exponents are
presented in \ref{Sec4}.
Sec. \ref{Sec5} brings a short recapitulation and concluding remarks.

\section{Model and its exactly solvable cases} \label{Sec2}

\subsection{Basic facts about the model}
\begin{figure}
\begin{center}
\includegraphics[width=0.45\textwidth,clip]{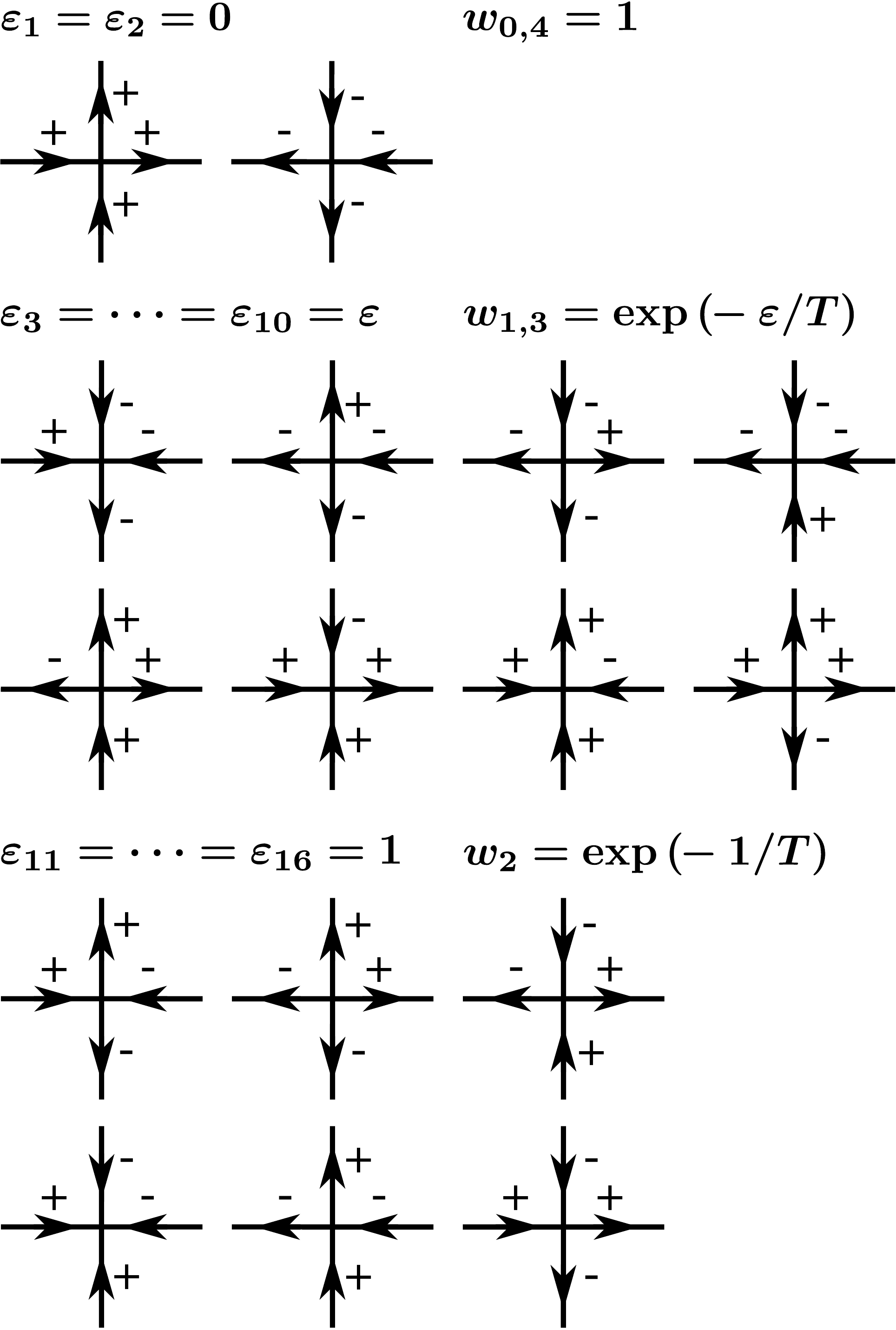}
\caption{Vertex weights of the symmetric 16-vertex model on the square lattice,
invariant with respect to the flip of all adjacent edge states 
$+\leftrightarrow -$.}
\label{fig:vertexweights}
\end{center}
\end{figure}

The general two-state vertex model on the square lattice of $N$ $(N\to\infty)$
sites is defined as follows.
Each lattice edge can be in one of two states. 
These states will be denoted either by $\pm$ signs or by ``dipole'' arrows: 
the right/up oriented arrow corresponds to the $(+)$ state, while 
the left/down arrow to the $(-)$ state.
With each vertex we associate the set of $2^4$ possible Boltzmann weights
$w(s_1,s_2,s_3,s_4) = \exp\left[ -\varepsilon(s_1,s_2,s_3,s_4)/T\right]$. 
In units of $k_{\rm B}=1$, both the energy $\varepsilon(s_1,s_2,s_3,s_4)$ 
and the temperature $T$ are taken as dimensionless. 
For the symmetric version of the model, the vertex weights are 
invariant with respect to any permutation of state variables 
$(s_1,s_2,s_3,s_4)$.
Let us denote by $w_i=\exp(-\varepsilon_i/T)$ $(i=0,1,\ldots,4)$ the vertex 
weight with $i$ incident edges in the $(-)$ state and the remaining $4-i$ 
incident edges in the $(+)$ state. 
Thus among the 16 possible configurations of vertex states there is 1
configuration corresponding to each of the vertex weights $w_0$ 
and $w_4$, 4 configurations corresponding to each of $w_1$ and $w_3$, 
and 6 configurations corresponding to $w_2$, see 
Fig.~\ref{fig:vertexweights}.

Thermal equilibrium of the system is determined by the (dimensionless) 
free energy per site
\begin{equation}
- \frac{f(\{ w\})}{T} = \lim_{N\to\infty} \frac{1}{N} \ln Z(\{ w\}) ,
\end{equation}
where
\begin{equation} \label{partition}
Z(\{ w\}) = \sum_{\{s\}} \prod_{\rm vertex} {\rm (weights)} ,
\end{equation} 
is the partition function with the summation going over all possible edge 
configurations and the product being over all vertex weights on 
the lattice.
The mean concentration $c_i$ of the vertices with weight $w_i$ is given by
\begin{equation} \label{ci}
c_i = - w_i \frac{\partial}{\partial w_i} \frac{f(\{ w\})}{T}  \qquad
(i=0,\ldots,4) .
\end{equation}
The mean concentrations are constrained by the obvious normalization
condition $\sum_{i=0}^4 c_i = 1$.
The mean-value of the edge-state variable
\begin{equation} \label{polarization}
P = \frac{1}{4} \sum_{i=0}^4 (4-2i) c_i
\end{equation}
defines the polarization.
When one applies an isotropic electric field $E$ (with the same strength
along either of the two axes), each arrow dipole $s=\pm 1$ 
acquires the energy $-E s$. 
Since every dipole belongs to just two vertices, the vertex weights are 
modified to
\begin{equation}
w_i(E) = w_i \exp\left[E(2-i)/T\right] .
\end{equation}
One can trivially extend the definitions of the vertex concentrations (\ref{ci})
and the polarization (\ref{polarization}) to $E\ne 0$, 
with the corresponding notations $c_i(E)$ and $P(E)$.
Then the polarization susceptibility reads as
\begin{equation} \label{chi}
\chi = \lim_{E\to 0} \frac{\partial P(E)}{\partial E} 
= \frac{1}{2} \sum_{i,j=0}^4 (2-i) (2-j) \chi_{ij} ,
\end{equation}
where the elements
\begin{equation}
\chi_{ij} = - \frac{\partial}{\partial\epsilon_j} c_i(E=0)
\end{equation}
form the tensor of generalized susceptibilities.

The partition function of the general two-state vertex model 
is invariant under the $O(2)$ gauge transformation of the
vertex weights \cite{Wegner73,Gaaff75}. 
On the square lattice with the coordination number 4, 
the gauge transformation reads as
\begin{eqnarray}
\tilde{w}(s_1,s_2,s_3,s_4) & = & \sum_{s'_1,s'_2,s'_3,s'_4} V_{s_1s'_1}(y) 
V_{s_2s'_2}(y) V_{s_3s'_3}(y) \nonumber \\ & & \times
V_{s_4s'_4}(y) w(s'_1,s'_2,s'_3,s'_4) .
\end{eqnarray}
Here, $V_{ss'}(y)$ are the elements of the matrix
\begin{equation}
{\bf V}(y) = \frac{1}{\sqrt{1+y^2}} 
\begin{pmatrix}
1 & y \\ y & -1
\end{pmatrix} 
\end{equation}
with rows (columns) indexed from up to down (left to right) as $+,-$
and a free (real) gauge parameter $y$.
For the symmetric version of the vertex model, the gauge transformation keeps 
the permutation symmetry of the vertex weights \cite{Wu89}, namely
\begin{eqnarray}
\tilde{w}_i & = & \sum_{j=0}^4 W_{ij}(y) w_j \quad (i=0,1,\ldots,4) , \\
W_{ij}(y) & = &  \frac{1}{(1+y^2)^2} \sum_{k=0}^{\min(i,j)} 
{i\choose k} {4-i\choose j-k} (-1)^k y^{i+j-2k} . \nonumber \\ & &
\end{eqnarray}

The points in the vertex-weight parameter space, which can be mapped 
onto themselves by gauge transformation with a nontrivial 
(point-dependent) value of $y\ne 0$, form the so-called self-dual manifold.
The self-dual manifold for the symmetric 16-vertex model is given by \cite{Wu89}
\begin{eqnarray} 
w_0^2 w_3 - w_1 w_4^2 - 3 w_2 (w_0-w_4)(w_1+w_3) & & \nonumber \\
+ (w_1-w_3) \left[ w_0 w_4 + 2 (w_1+w_3)^2 \right] & = & 0 . \label{selfdual} 
\end{eqnarray}
Its importance consists in the fact that all critical points of 
the second-order phase transitions are confined to this subspace 
of the vertex weights.

In this work, we restrict ourselves to the symmetric 16-vertex model whose
vertex weights are invariant with respect to the flip of all adjacent edge 
states $(+)\leftrightarrow (-)$.
The vertex weights are parametrized as follows
\begin{equation} \label{spinflip}
w_0=w_4=1 , \quad w_1=w_3=e^{-\epsilon/T} , \quad w_2= e^{-1/T} ,
\end{equation}
see also Fig.~\ref{fig:vertexweights}, where the real energy parameter 
$\epsilon\ge 0$. 
It can be checked that this choice of vertex weights automatically 
satisfies the self-dual condition (\ref{selfdual}). 
Thus, for a fixed value of the energy $\varepsilon$, there should exist 
a critical temperature $T_c$ at which the second-order phase transition
takes place.
The order parameter is always the mean polarization $P$, see 
Eq. (\ref{polarization}). 
In the disordered phase, for $T>T_c$, the state-flip symmetry 
of vertex weights implies the equality of mean vertex 
concentrations $c_i=c_{4-i}$ $(i=0,1)$ and $P$ vanishes.
In the ordered phase, for $T<T_c$, the state-flip symmetry breaking causes
that $c_i\ne c_{4-i}$ and the spontaneous polarization $P$ becomes nonzero.
At $T_c$, $P$ is nonanalytic in $T_c-T$:
\begin{equation} \label{spontaneousmagn}
P \propto (T_c-T)^{\beta_{\rm e}} , \qquad T\to T_c^- 
\end{equation}
with $\beta_{\rm e}$ (the subscript e means ``electric'') being 
the critical exponent. 
If a small isotropic external electric field $E$ is applied to the
vertex system just at the critical temperature, the polarization behaves as
\begin{equation} \label{critfield}
P(E) \propto E^{1/\delta_{\rm e}} , \qquad T=T_c ,
\end{equation}
where $\delta_{\rm e}$ is another critical exponent.
Close to the critical point, the polarization susceptibility (\ref{chi}) 
exhibits a singularity of type
\begin{equation} \label{susceptibility}
\chi \propto \frac{1}{\vert T_c-T\vert^{\gamma_{\rm e}}} ,
\end{equation}
where the critical exponent $\gamma_{\rm e}$ is assumed to be the same for
both limits $T\to T_c^-$ and $T\to T_c^+$.
The pair arrow-arrow correlation function exhibits the large-distance behavior
\begin{equation}
G_{\rm e}(r) \propto \frac{1}{r^{\eta_{\rm e}}} \exp(-r/\xi) ,
\qquad r\to \infty .
\end{equation} 
Approaching the critical point, the correlation length $\xi$
diverges as
\begin{equation}
\xi \propto \frac{1}{\vert T_c-T\vert^{\nu_{\rm e}}} . 
\end{equation} 
The divergence of $\xi$ at $T=T_c$ reflects the fact that the short-range
(exponential) decay changes into the long-range (inverse power-law) 
decay at $T=T_c$, which is characterized by the critical exponent 
$\eta_{\rm e}$.

Having at one's disposal the two critical exponents $\beta_{\rm e}$ and 
$\delta_{\rm e}$, the remaining ones (considered in this work) can be 
calculated by the 2D scaling relations \cite{Baxterbook}:
\begin{subequations}
\begin{eqnarray}
\gamma_{\rm e} & = & \beta_{\rm e} \left( \delta_{\rm e} -1 \right) , 
\label{scalinga} \\
\nu_{\rm e} & = & \frac{1}{2} \beta_{\rm e} \left( \delta_{\rm e} +1 \right) , 
\label{scalingb} \\
\eta_{\rm e} & = & \frac{4}{\delta_{\rm e}+1} . \label{scalingc}
\end{eqnarray}
\end{subequations}

\subsection{Ising point}
The symmetric 16-vertex model can be mapped onto the Ising model on
the square lattice under the vertex-weight constraint
\cite{Samaj91,Samaj92}
\begin{equation} \label{mappingIsing}
w_0 w_2 w_4 - w_0 w_3^2 - w_1^2 w_4 +2 w_1 w_2 w_3 -w_2^3 = 0 .
\end{equation}
For the state-flip symmetry of the vertex weights (\ref{spinflip}), 
this equation takes the form
\begin{equation} \label{Eq1}
1 + e^{1/T} = 2 e^{2(1-\varepsilon)/T} .
\end{equation}
As concerns the parameters of the Ising model for the state-flip symmetry, 
the external magnetic field acting on spins $H=0$ and the (dimensionless) 
coupling $J$ between the nearest-neighbor spins is given by
\begin{equation} \label{Eq2}
J = \frac{1}{2} \ln\left( \frac{w_1}{w_2} \right) =
\frac{1-\varepsilon}{2 T} .
\end{equation} 
The known critical value of the Ising coupling reads \cite{Baxterbook}
\begin{equation}
J_c = \frac{1}{2} \ln\left( 1+\sqrt{2} \right) .
\end{equation}
Consequently, Eqs. (\ref{Eq1}) and (\ref{Eq2}) imply the following critical 
parameters of the symmetric 16-vertex model:
\begin{eqnarray}
\varepsilon^{\rm (I)} & = & 1 - \frac{\ln(1+\sqrt{2})}{\ln(5+4\sqrt{2})} 
= 0.627516\ldots , \label{Isingenergy} \\ 
T^{\rm (I)}_c & = & \frac{1}{\ln(5+4\sqrt{2})} = 0.422618\ldots . 
\label{Isingtemperature}
\end{eqnarray}

In contrast to standard mappings of models on dual lattices, the mapping 
between the symmetric 16-vertex and the Ising models is made on 
the {\em same} square lattice \cite{Samaj91,Samaj92}.
The relation between the polarization of the symmetric 16-vertex model
and the magnetization of the equivalent Ising system can be derived 
by using the technique presented in Ref. \cite{Kolesik91}.
This relation is linear and, therefore, the critical exponents of the
symmetric 16-vertex model are identical to the ones of the Ising model.
The Ising critical exponents are summarized in Table \ref{tab}.

\begin{table}[htb] 
\begin{center}
\begin{tabular}{|l|c|c|c|c|c|} \hline
exponent & $\beta_{\rm e}$ & $\delta_{\rm e}$ & $\gamma_{\rm e}$ 
& $\nu_{\rm e}$ & $\eta_{\rm e}$ \\ \hline \hline
$\varepsilon^{\rm (I)}\approx 0.6275$ & $1/8$ & $15$ & $7/4$ & $1$ & $1/4$ \\
\hline
$\varepsilon\to\infty$ & $1/8$ & $11$ & $5/4$ & $3/4$ & $1/3$ \\
\hline
\end{tabular}
\end{center}
\caption{List of electric critical exponents for the symmetric
16-vertex model at the exactly solvable Ising and the Baxter 8-vertex points.}
\label{tab}
\end{table}

\subsection{8-vertex point}
When $\varepsilon\to\infty$, the vertex weights $w_1$ and $w_3$, 
corresponding in Fig.~\ref{fig:vertexweights} to configurations with 
odd numbers of $(+)$, or equivalently $(-)$, edge states, vanish.
The consequent Baxter's 8-vertex model has vertex-weight parameters 
$a=w_0=w_4=1$ and $b=c=d=w_2=\exp(-1/T)$ \cite{Baxterbook}.
The vertex system exhibits the ferroelectric-$A$ phase for $a>b+c+d$.
The second-order transition between the ferroelectric-A and disordered phases 
takes place at
\begin{equation} \label{Baxtertemp}
a_c=b_c+c_c+d_c , \qquad T_c = \frac{1}{\ln 3} = 0.910239\ldots .
\end{equation}
Introducing the auxiliary parameter
\begin{equation}
\mu = 2 \arctan \left( \sqrt{\frac{a_cb_c}{c_cd_c}} \right) 
= \frac{2\pi}{3} ,
\end{equation}
according to Ref. \cite{Krcmar18} the electric critical exponents
are given by 
\begin{eqnarray}
\beta_{\rm e} & = & \frac{\pi-\mu}{4\mu} = \frac{1}{8} , \nonumber \\
\delta_{\rm e} & = & \frac{3\pi+\mu}{\pi-\mu} = 11 , \nonumber \\
\gamma_{\rm e} & = & \frac{\pi+\mu}{2\mu} = \frac{5}{4} ,  \\
\nu_{\rm e} & = & \frac{\pi}{2\mu} = \frac{3}{4} , \nonumber \\
\eta_{\rm e} & = & 1-\frac{\mu}{\pi} = \frac{1}{3} . \nonumber
\end{eqnarray}
These critical exponents are listed in Table \ref{tab}.

\section{Numerical method} \label{Sec3}

\begin{figure}
\centering
\includegraphics[width=0.4\textwidth,clip]{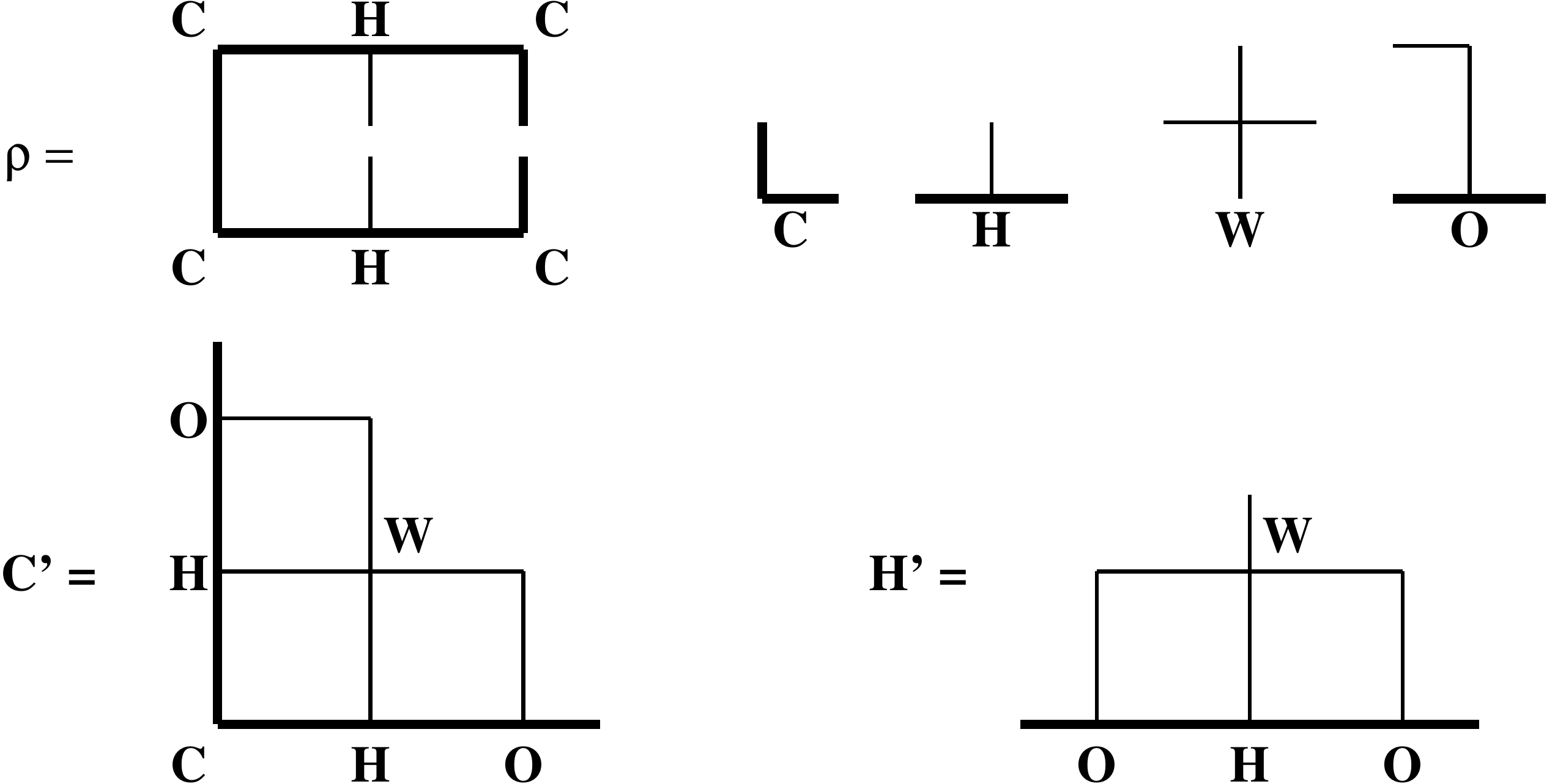}
\caption{The CTMRG renormalization process. 
The density matrix $\rho$ is composed of four transfer matrices $C$. 
The expansion process of the corner transfer matrix 
$C\to C'= O^{\dagger} HWCHO$ and the half-row transfer matrix 
$H\to H' = O^{\dagger} HWO$ from the previous iteration RG Step, 
see the text.}
\label{fig:ctmrg}
\end{figure}

The CTMRG method \cite{Nishino96,Nishino97,Ueda05} is based on Baxter's
technique of corner transfer matrices \cite{Baxterbook}. 
Each quadrant of the square lattice with size $L\times L$ is represented 
by the corner transfer matrix $C$. 
The reduced density matrix is defined by $\rho = \Tr' C^4$
(where the partial trace $\Tr'$ is taken), so that the partition function
$Z = \Tr\rho$, see Fig.~\ref{fig:ctmrg}. 
The number of degrees of freedom grows exponentially with $L$ and 
the density matrix is used in the process of their reduction. 
Namely, degrees of freedom are iteratively projected to the space generated 
by the eigenvectors of the reduced density matrix $\rho$
with the largest eigenvalues. 
The projector on this reduced space of dimension $m$ is denoted by $O$; 
the larger the truncation parameter $m$ is taken, 
the better precision of the results is attained. 
In each iteration the linear system size is expanded from $2L$ to $2L+2$
via the inclusion of the Boltzmann weight $W$ of the basic vertex 
(see Fig.~\ref{fig:vertexweights}).
The expansion process transforms the corner transfer matrix $C$ to $C'$
and the half-row transfer matrix $H$ to $H'$ in the way represented 
schematically in Fig.~\ref{fig:ctmrg}.
The thin (thick) lines represent renormalized (multi-) arrow variables 
obtained after the renormalization. 
The fixed boundary conditions are imposed, i.e., the state $(-)$
is fixed on the boundary arrows only.
This choice ensures a quicker convergence of the method in the thermodynamic 
limit.

\section{Numerical results} \label{Sec4}

\begin{figure}
\begin{center}
\includegraphics[width=0.45\textwidth,clip]{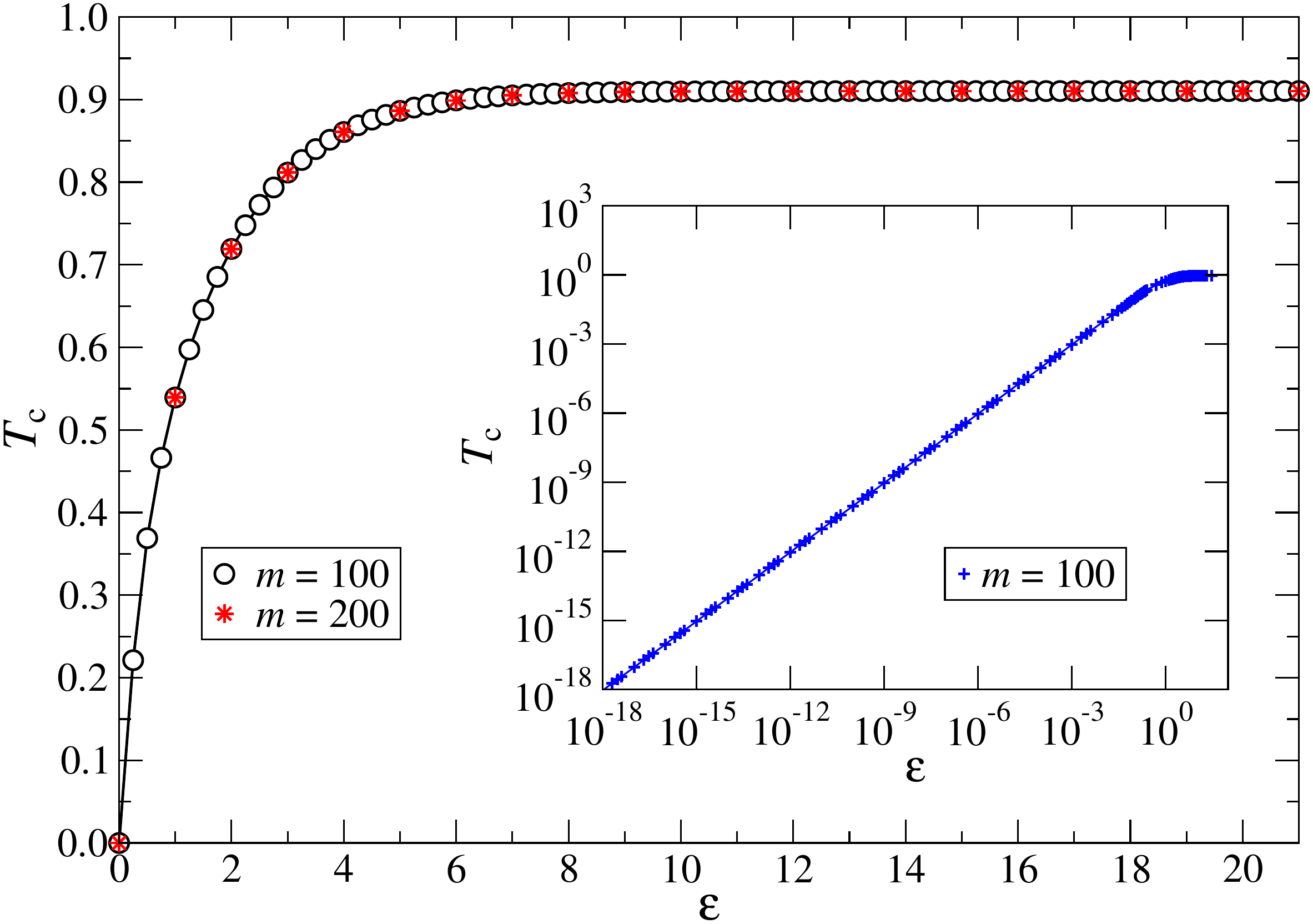}
\caption{The $\varepsilon$-dependence of the critical temperature $T_c$
of the symmetric 16-vertex model, for dimension of the truncated space 
$m=100$ (open circles) and $m=200$ (open circles with stars).  
The inset shows a linear dependence of $T_c(\varepsilon)$ for small values
of $\varepsilon$.}
\label{fig:criticaltemperature}
\end{center}
\end{figure}

According to Eq. (\ref{spontaneousmagn}), the critical temperature $T_c$
is the lowest temperature at which $P=0$ or, equivalently, the highest
temperature at which $P\ne 0$.
Based on comparison with the known values of the Ising (\ref{Isingtemperature})
and Baxter's (\ref{Baxtertemp}) critical temperatures, the error in estimation 
of $T_c(\varepsilon)$ is of order $10^{-4}$ for all values of $\varepsilon$. 
The error is even smaller (of order $10^{-5}$) when fitting data
for the spontaneous polarization close to the critical point 
according to (\ref{spontaneousmagn}).
Numerical results for the $\varepsilon$-dependence of the critical
temperature are shown in Fig.~\ref{fig:criticaltemperature}.
We see that $T_c(\varepsilon)$ is only weakly affected by dimension
of the truncated space $m=100$ and $m=200$, which means that our
results reached the sufficient accuracy.

The inset of Fig.~\ref{fig:criticaltemperature} documents the log-log 
plot of the small-$\varepsilon$ behavior of $T_c(\varepsilon)$.  
The power-law least-square fitting at low $\varepsilon<10^{-8}$ yields 
\begin{equation}
T_c(\varepsilon) = -6.6\times 10^{-18} + 0.954(5) \varepsilon^{0.9998(3)} ,
\end{equation}
where the absolute term is on the accuracy border of the computer
(the machine precision).
We conclude that in the limit of small $\varepsilon$ 
the critical temperature converges to zero linearly.
On the other hand, as $\varepsilon$ increases, the critical temperature
saturates quickly to the value $0.91024$ which is close to the asymptotic
$\varepsilon\to\infty$ analytic result (\ref{Baxtertemp}) of 
the 8-vertex model.

\begin{figure}
\begin{center}
\includegraphics[width=0.45\textwidth,clip]{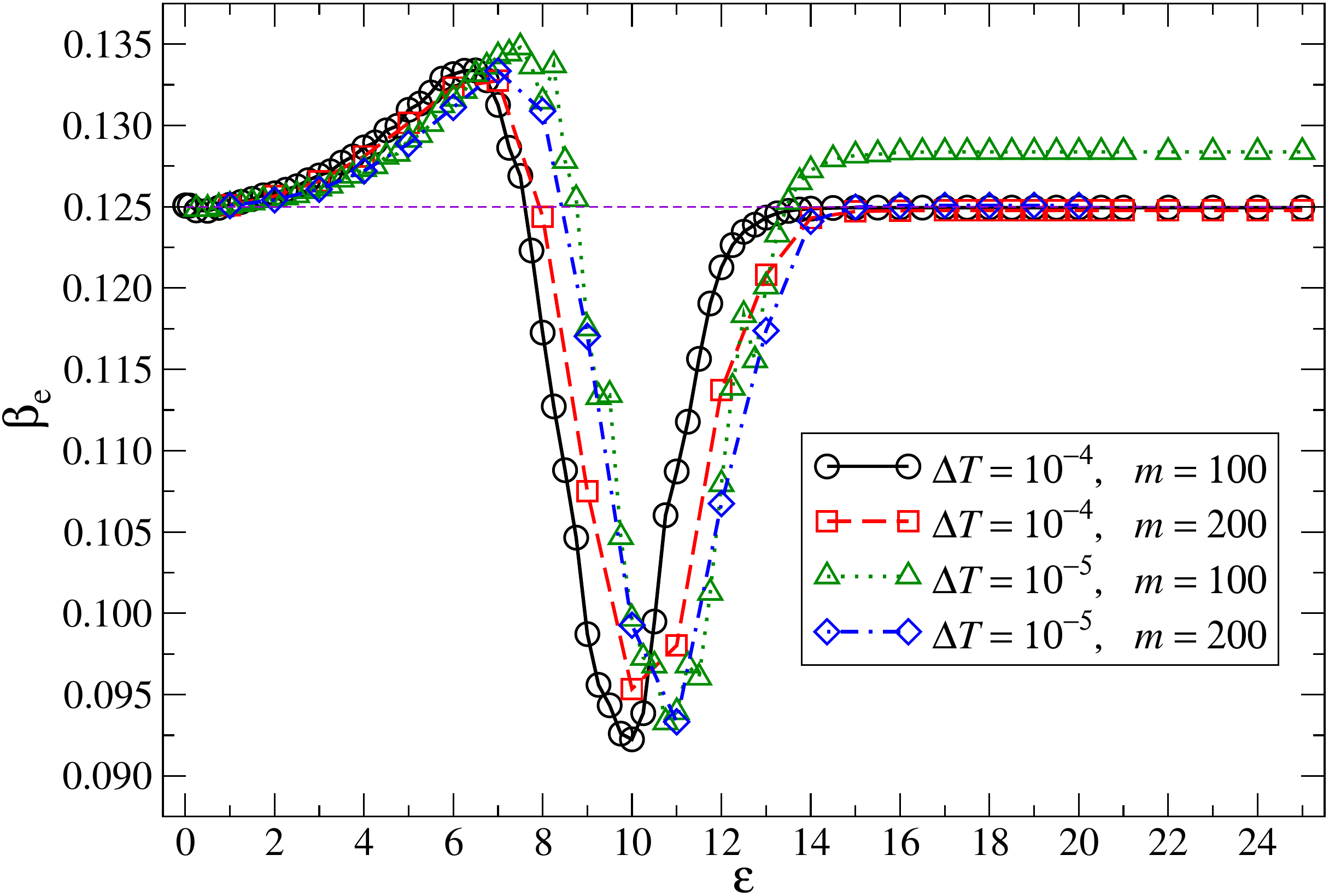}
\caption{The $\varepsilon$-dependence of the critical exponent $\beta_{\rm e}$
for the symmetric 16-vertex model with the temperature steps 
$\Delta T = 10^{-4}$ and $10^{-5}$ and dimensions of the truncated space
$m=100$ and $m=200$.}
\label{fig:betae}
\end{center}
\end{figure}

The critical exponent $\beta_{\rm e}$ is expected to interpolate 
between the same values $1/8$ at small and large $\varepsilon$. 
It is calculated by fitting the polarization data according to formula 
(\ref{spontaneousmagn}).
With $T_c$ fixed in the previous calculation, we have selected a series of 
temperatures below the threshold value $T_c-0.0002$ with 
a temperature spacing (discretization step) $\Delta T$ at which 
the polarization is evaluated.
For each value of $\varepsilon$, we have generated 6 polarization values with
$\Delta T = 10^{-4}$ and 30 polarization values with $\Delta T = 10^{-5}$
if taking dimension of the truncated space $m=100$ and $m=200$.
The corresponding $\varepsilon$-dependences of the critical exponent 
$\beta_{\rm e}$ within the range of $\varepsilon \in [0,25]$, are pictured in 
Fig.~\ref{fig:betae}. 
We see that the too small value of the temperature step $\Delta T = 10^{-5}$ 
and $m=100$ (triangles) leads in the region of large $\varepsilon$ 
incorrectly to $\beta_{\rm e}>1/8$.
If increasing the accuracy to $m=200$ (diamonds), data converge
to the correct value $\beta_{\rm e}=1/8$ at large $\varepsilon$.
On the other hand, for a larger temperature step $\Delta T = 10^{-4}$, 
both $m=100$ (circles) and $m=200$ (squares) data are consistent 
with $\beta_{\rm e}=1/8$ at large $\varepsilon$.
We refer to the parameters $\Delta T = 10^{-4}$ and $m=200$ as 
the optimal ones.
The choice of these optimal parameters correctly reproduces the exact results 
for the Ising $\varepsilon^{\rm (I)}\approx 0.6275$ and the 8-vertex
$\varepsilon\to\infty$ models and, therefore, it is expected to be
adequate also in the transition region between the two solvable cases.  
The above-discussed cases are presented in Fig.~\ref{fig:betae} to judge 
the relative accuracy of the relevant data in the transition region of 
$\varepsilon$ values.
In the interval of $\varepsilon\lesssim 2$ containing the Ising
point $\varepsilon=0.627516\ldots$, the exponent is roughly constant
$\beta_{\rm e}=1/8$.
In the transition region
$2\lesssim \varepsilon \lesssim 14$, $\beta_{\rm e}$ varies nonmonotonously 
as a function of $\varepsilon$.
For $\varepsilon\gtrsim 14$, the exponent $\beta_{\rm e}$ is again 
constant and acquires its Baxter's $(\varepsilon\to\infty)$
value $\beta_{\rm e}=1/8$, as it should be.

\begin{figure}
\begin{center}
\includegraphics[width=0.45\textwidth,clip]{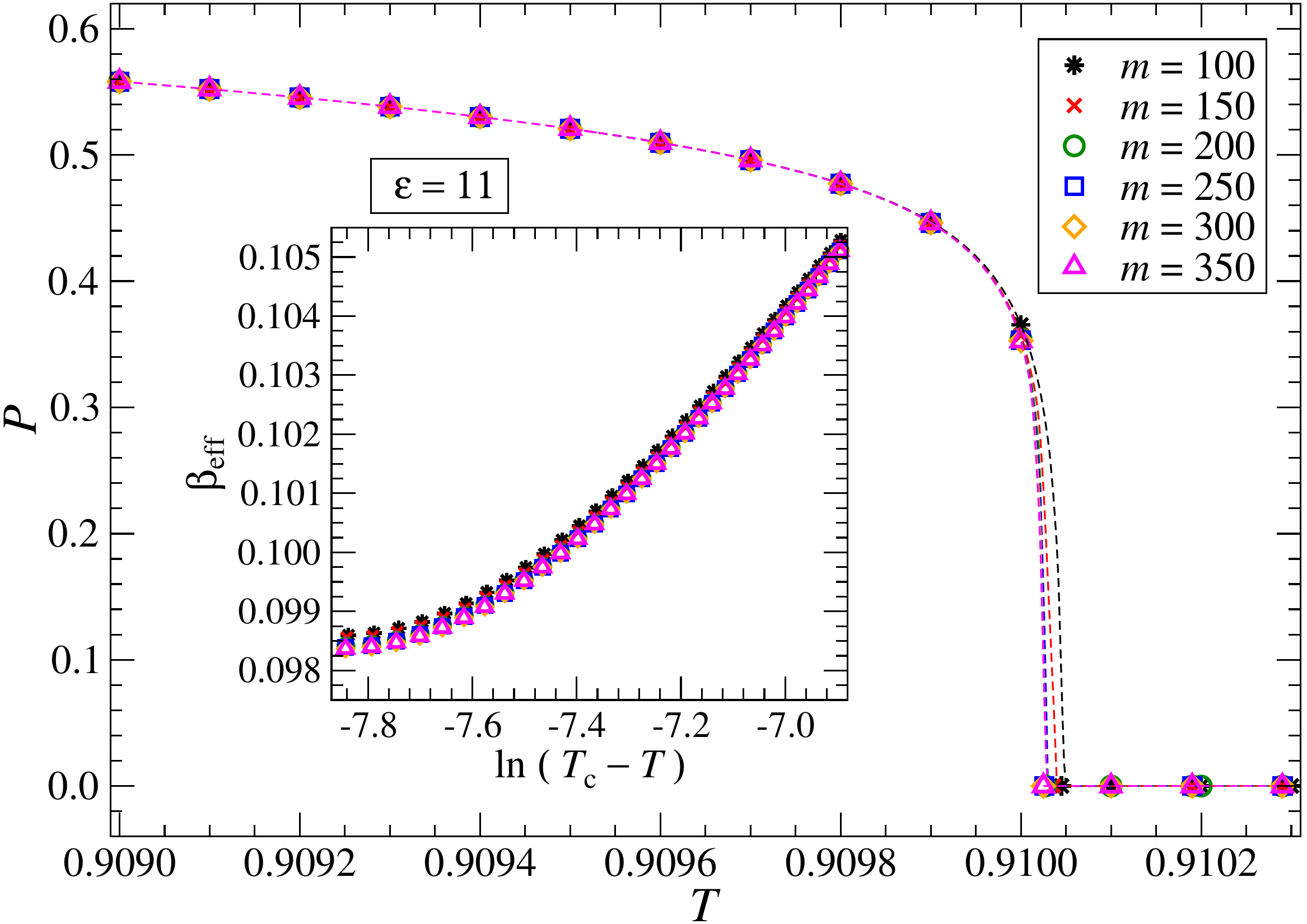}
\caption{The polarization (the order parameter) $P$ as a function of 
temperature $T$ calculated at $\varepsilon=11$ and for various numbers of 
the states reduction $m$ ranging from 100 to 350. 
The inset shows the dependence of the effective critical exponent 
$\beta_{\rm eff}$ on the logarithmic distance of the temperature from 
the critical temperature $T_c$; as $T$ approaches $T_c$ from below, 
$\beta_{\rm eff}(T \to T_c) \to \beta_{\rm e}$.}
\label{fig:testbeta}
\end{center}
\end{figure}

To document the accuracy of the CTMRG method, we present in 
Fig. \ref{fig:testbeta} the evaluation of the critical exponent $\beta_{\rm e}$ 
for the energy $\varepsilon=11$ which lies in the transition region.
The truncation orders $m$ range from 100 to 350 and the optimal
$\Delta T = 10^{-4}$ is chosen.
We define the effective exponent $\beta_{\rm eff}$ as follows 
\begin{equation}
P(T) \propto (T_c - T)^{\beta_{\rm eff}(T)} , \qquad 
\mbox{for $0 \ll T \leq T_c$,}
\end{equation}
where the prefactor does not depend on the temperature.
As $T$ approaches $T_c$ from below, the exponent $\beta_{\rm eff}(T)$ converges 
to the electric exponent $\beta_{\rm e}$ we are looking for:
\begin{equation}
\lim_{T \to T_c} \beta_{\rm eff}(T) 
= \lim_{T \to T_c} \frac{\partial \ln P(T)}{\partial \ln(T_c- T)} 
= \beta_{\rm e} .
\end{equation}
As seen in the inset of Fig. \ref{fig:testbeta}, the data for 
$\beta_{\rm eff}$ as the function of the logarithmic distance of 
the temperature from $T_c$ get converged starting from $m=200$.
The plot of $\beta_{\rm eff}$ is almost constant for $\ln(T_c-T)<-7.8$;
with regard to the fine scale on the $\beta_{\rm eff}$-axis this fact
permits an accurate determination of $\beta_{\rm e}$.
Since the accuracy of the CTMRG method is superior to standard numerical
approaches like Monte Carlo simulations, the continuous variation of 
$\beta_{\rm e}$ in Fig. \ref{fig:betae}, ranging in the large interval 
$2\lesssim \varepsilon \lesssim 14$, exhibits a relatively large amplitude
exceeding by orders the error bars in the exponent determination
by the present technique. In the same manner, we have analyzed the
critical exponents investigated in the remaining part of the paper.

\begin{figure}
\begin{center}
\includegraphics[width=0.45\textwidth,clip]{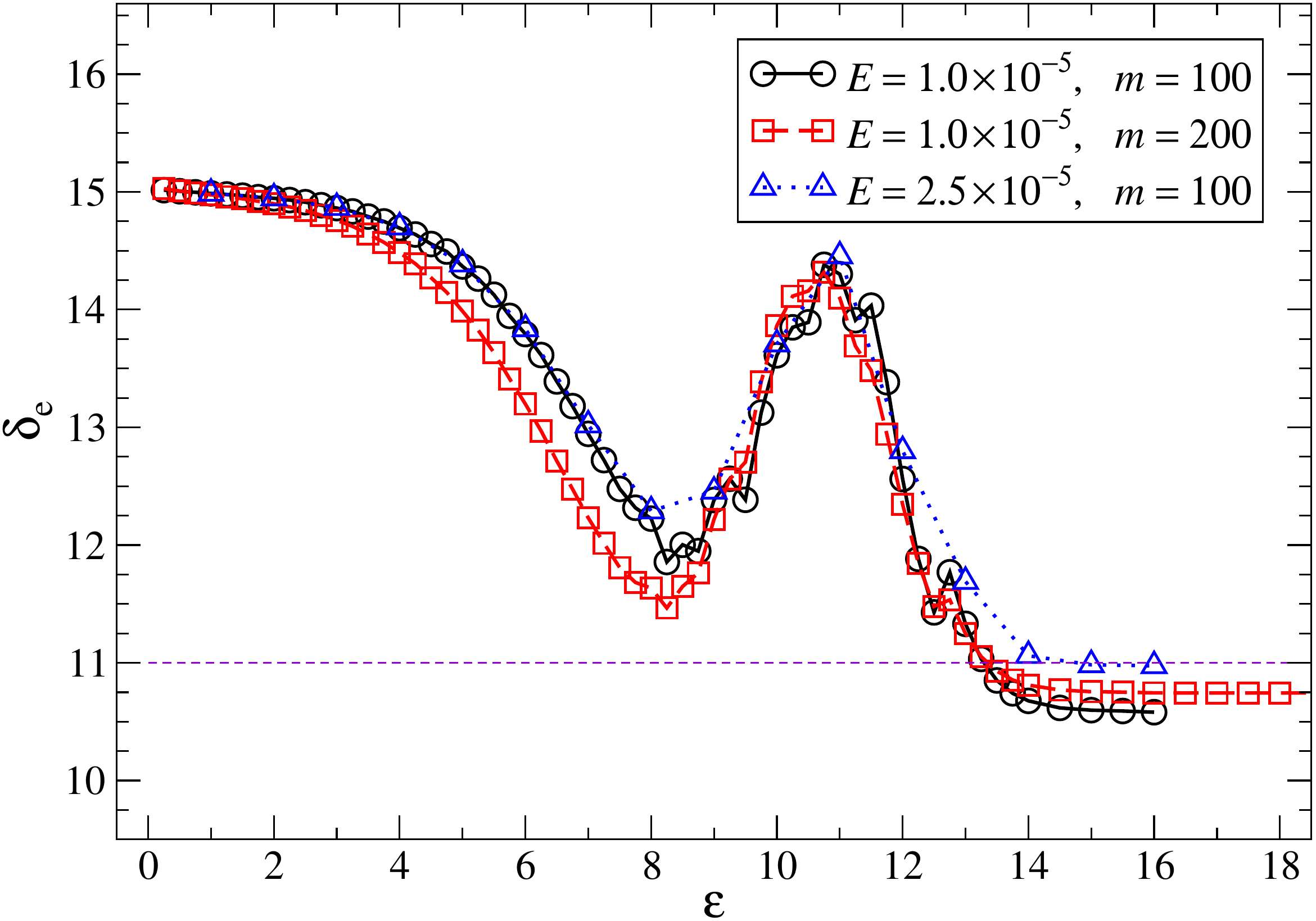}
\caption{The $\varepsilon$-dependence of the critical exponent $\delta_{\rm e}$
for the symmetric 16-vertex model.
Data are generated for the electric field $E=10^{-5}$ at approximation orders
$m=100$ (circles) and $m=200$ (squares), and the optimal
$E=2.5\times 10^{-5}$ at $m=100$ (triangles).}
\label{fig:deltae}
\end{center}
\end{figure}

As seen in Table \ref{tab}, the critical exponent $\delta_{\rm e}$ 
is expected to interpolate between the values $15$ at small $\varepsilon$ 
and $11$ at large $\varepsilon$. 
It is calculated by fitting the polarization data at the critical 
temperature $T_c$ according to the relation (\ref{critfield}) 
which can be rewritten as
\begin{equation} \label{fitdelta}
\delta_{\rm e} = \mathop{\lim}_{E\to 0}
\left( \frac{\partial \ln P}{\partial \ln E} \right)^{-1} .
\end{equation}
This formula has to be considered for a very small value of field $E$,
but not too small to avoid numerical errors due to the critical state
of the vertex system.
The obtained data for $E=10^{-5}$ and $2.5\times 10^{-5}$ are presented in 
Fig.~\ref{fig:deltae}, within the range of $\varepsilon \in [0,18]$.
Data for $E=10^{-5}$, evaluated at approximation orders 
$m=100$ (circles) and $m=200$ (squares), converge below 
the anticipated value $11$.
On the other hand, numerical data for the optimal field $E=2.5\times 10^{-5}$ 
evaluated at approximation order $m=100$ (triangles) lie close to 
the previous data for $E=10^{-5}$ with $m=200$ in the region
$0\lesssim \varepsilon \lesssim 12$ and tend to the correct value $11$
for large values of $\varepsilon$.

\begin{figure}
\begin{center}
\includegraphics[width=0.45\textwidth,clip]{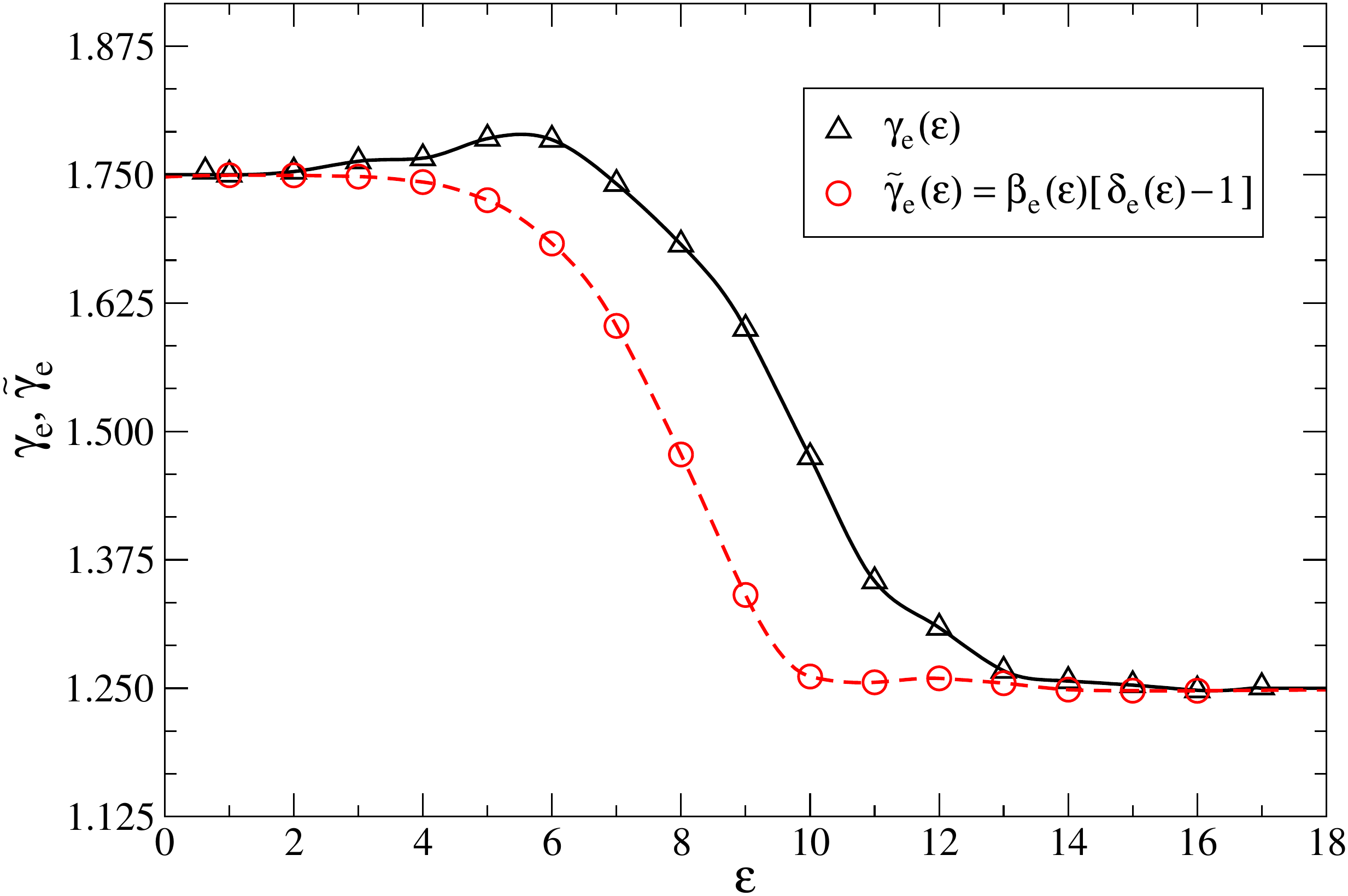}
\caption{The $\varepsilon$-dependence of the critical exponent $\gamma_{\rm e}$
for the symmetric 16-vertex model.
Data are generated from fitting of the formula (\ref{susceptibility}), 
in the region $T>T_c$ and the susceptibility values $\chi\in[10000,50000]$ 
calculated with dimension of the truncated space $m=100$ (triangles). 
The exponent $\tilde{\gamma}_{\rm e}$, calculated by inserting the previous 
data for $\beta_{\rm e}$ and $\delta_{\rm e}$ into the scaling relation 
(\ref{scalinga}), is represented by circles.} 
\label{fig:gammae}
\end{center}
\end{figure}

The critical exponent $\gamma_{\rm e}$ is expected to interpolate between 
$7/4$ at small $\varepsilon$ and $5/4$ at large values of $\varepsilon$. 
This exponent is calculated by fitting the susceptibility data
according to the formula (\ref{susceptibility}). 
The fitting is performed in the region $T>T_c$ with the susceptibility
functional values from the interval $\chi\in[10000,50000]$.
Within the range of $\varepsilon \in [0,18]$, the obtained $m=100$ data
are represented by triangles in Fig.~\ref{fig:gammae}.
Data tend for small and large values of $\varepsilon$ correctly to 
$7/4$ and $5/4$, respectively.
Because the fits of the singular formula (\ref{susceptibility}) are 
accompanied by relatively large errors, we have calculated alternatively 
$\tilde{\gamma}_{\rm e}$ by inserting the previous data for $\beta_{\rm e}$
and $\delta_{\rm e}$ into the scaling relation (\ref{scalinga}).
Hereinafter, we adopt convention that an exponent deduced by using scaling 
relations will be denoted by a tilde on its top. 
The data for $\tilde{\gamma}_{\rm e}$ are represented in 
Fig.~\ref{fig:gammae} by circles.   
Note that the plot exhibits a monotonous decay.

\begin{figure}
\begin{center}
\includegraphics[width=0.45\textwidth,clip]{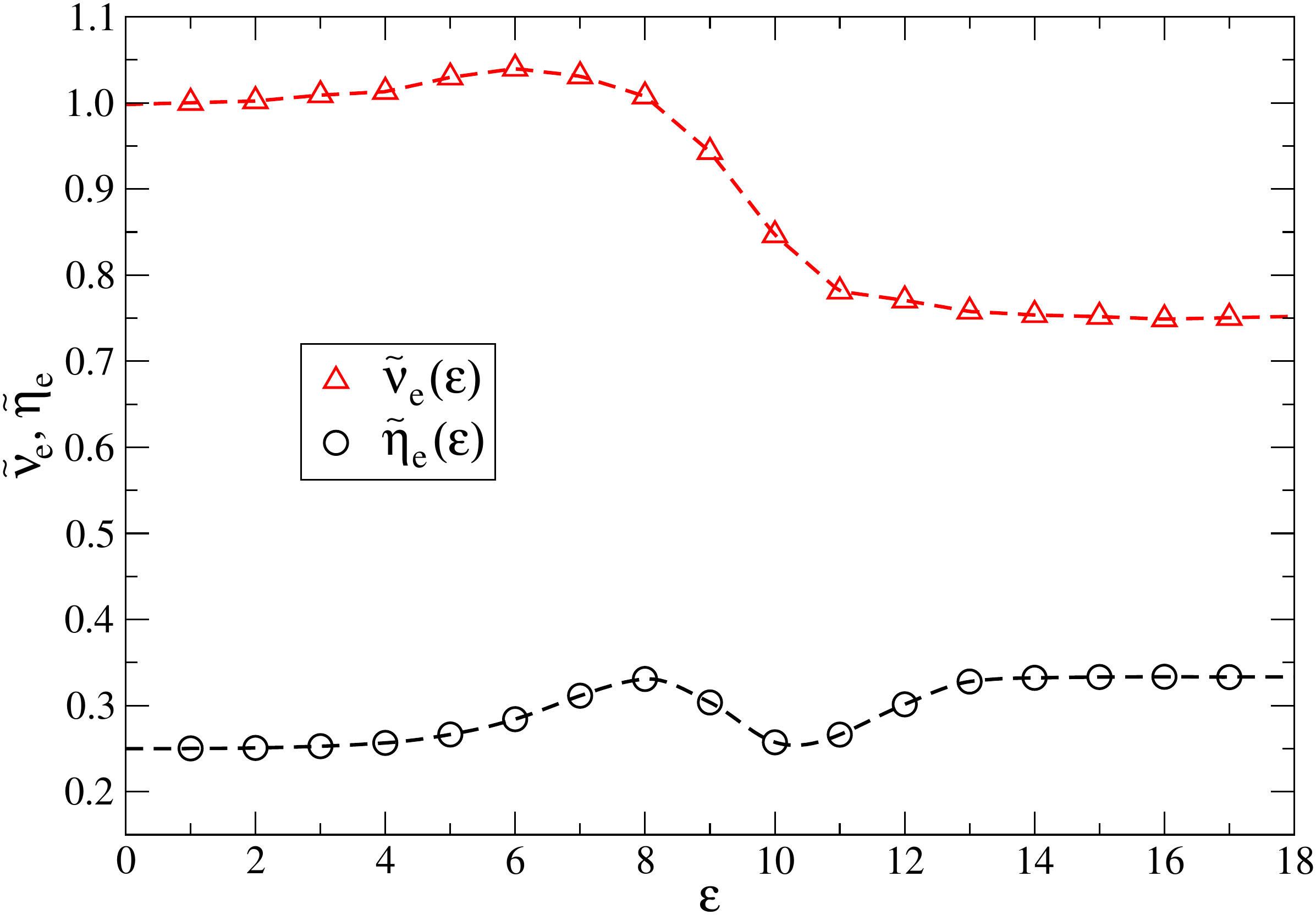}
\caption{The critical exponents $\tilde{\nu}_{\rm e}$ (triangles)
and $\tilde{\eta}_{\rm e}$ (circles), calculated by inserting 
the previous data for $\beta_{\rm e}$ ($\Delta T = 10^{-4}$ and $m=200$) 
and $\delta_{\rm e}$ ($E=2.5\times 10^{-5}$ and $m=100$) into the second 
and third of scaling relations (\ref{scalingb}), respectively, 
as functions of $\varepsilon\in [0,18]$.}
\label{fig:nue}
\end{center}
\end{figure}

The critical exponents $\tilde{\nu}_{\rm e}$ and $\tilde{\eta}_{\rm e}$, 
calculated by inserting the previous data for $\beta_{\rm e}$ 
($\Delta T = 10^{-4}$ and $m=200$, squares in Fig.~\ref{fig:betae}) and 
$\delta_{\rm e}$ ($E=2.5\times 10^{-5}$ and $m=100$, triangles in
Fig.~\ref{fig:deltae}) into the scaling relations (\ref{scalingb}) and
(\ref{scalingc}), respectively, are represented as functions of $\varepsilon$ 
in Fig.~\ref{fig:nue} by triangles and circles.
Both plots exhibit nonmonotonous behavior.
The exponent $\tilde{\nu}_{\rm e}$ interpolates correctly between $1$
at small $\varepsilon$ and $3/4$ at large $\varepsilon$
and $\tilde{\eta}_{\rm e}$ interpolates correctly between $1/4$
at small $\varepsilon$ and $1/3$ at large $\varepsilon$.

The accuracy of the CTMRG method is superior to that of the standard 
numerical transfer matrix and Monte Carlo methods. 
The crucial feature of the present method is the extremely small error of 
order $10^{-4}-10^{-5}$ in the determination of the critical temperature 
$T_c(\varepsilon)$, whereas the error decreases to the machine precision
$(10^{-16})$ off $T_c(\varepsilon)$.
Having the precise value of the critical temperature, the fitting of the
critical exponents $\beta_{\rm e}$ by using (\ref{spontaneousmagn})
and $\gamma_{\rm e}$ by using (\ref{fitdelta}) is very accurate.
For the purpose of benchmarking, we employ a numerical method, the
Higher-Order Tensor Renormalization Group (HOTRG) \cite{HOTRG}, 
in order to provide an independent comparison with the CTMRG. 
We chose the HOTRG method for its numerical reliability and high accuracy 
with respect to the Monte Carlo simulations.
Having defined the absolute errors
${\cal E}_{T_{\rm c}}(\varepsilon) = \vert T_c^{\rm HOTRG}(\varepsilon)- 
T_c^{\rm CTMRG}(\varepsilon)\vert$ and ${\cal E}_{\beta_e}(\varepsilon)=\vert
\beta_{\rm e}^{\rm HOTRG}(\varepsilon)-\beta_{\rm e}^{\rm CTMRG}(\varepsilon) \vert$, 
we confirmed an excellent agreement between the CTMRG and HOTRG methods. 
In particular, we evaluated the errors at four points 
$\varepsilon = 9, 10, 11, 12$ of the transition region, where the exponents 
change rapidly, see Tab.~\ref{tab2}.

\begin{table}[htb] 
\begin{center}
\begin{tabular}{|c|c|c|} \hline
$\varepsilon$ & ${\cal E}_{T_{\rm c}}(\varepsilon)$
	      & ${\cal E}_{\beta_e}  (\varepsilon)$ \\ \hline \hline
 $9$ & $5 \times 10^{-6} $ & $6.8 \times 10^{-4}$ \\ \hline
$10$ & $3 \times 10^{-6} $ & $7.8 \times 10^{-4}$ \\ \hline
$11$ & $2 \times 10^{-6} $ & $2.8 \times 10^{-4}$ \\ \hline
$12$ & $2 \times 10^{-6} $ & $1.9 \times 10^{-3}$ \\ \hline
\end{tabular}
\end{center}
\caption{The absolute errors of the results for the critical temperatures 
$T_{\rm c}$ and the critical exponents $\beta_e$ obtained by using 
the CTMRG and HOTRG methods.}
\label{tab2}
\end{table}

Our first aim was to confirm that there is a line of critical 
points connecting the two exactly solvable Ising 
$\varepsilon^{\rm (I)}\approx 0.6275$ and 8-vertex $\varepsilon\to\infty$ 
points which belong to two different universality classes.
The order parameter, namely the polarization (\ref{polarization}),
is unique for all values of $\varepsilon\ge 0$.
The next question was whether the critical exponents are changing along the
line continuously, or they are constant in the regions of small and large
$\varepsilon$ with a discontinuous change at intermediate 
values of $\varepsilon$.
As seen in Figs.~\ref{fig:betae} and \ref{fig:deltae}, the variation 
of the two crucial critical exponents $\beta_{\rm e}$ and $\gamma_{\rm e}$ 
is considerable and takes place on a relatively large interval 
$2\lesssim \varepsilon \lesssim 14$.
With regard to the high accuracy of the CTMRG method, this fact supports
the scenario of a continuous change of the critical exponents along the line.
The same arguments hold as to the variation of the critical exponents 
$\gamma(\varepsilon)$ in Fig. \ref{fig:gammae} and $\eta(\varepsilon)$ 
in Fig. \ref{fig:nue}, but the variation of $\nu_{\rm e}$ in Fig. \ref{fig:nue}
permits the scenario of two universality classes only.

To judge the validity of the hypothesis of weak universality, it is
sufficient to test the thermal renormalized exponents $\beta_{\rm e}/\nu_{\rm e}$, 
$\gamma_{\rm e}/\nu_{\rm e}$ and the exponents $\delta_{\rm e}, \eta_{\rm e}$, 
which are independent of $\varepsilon$ if weak universality applies, 
at the two exactly solvable points.
In particular, from Table \ref{tab} we have
\begin{eqnarray}
\frac{\beta_{\rm e}}{\nu_{\rm e}} & = & \left\{
\begin{array}{lll}
\frac{1}{8} & & \varepsilon^{\rm (I)}\approx 0.6275 , \cr
\frac{1}{6} & & \varepsilon\to\infty ,
\end{array} \right. \\
\frac{\gamma_{\rm e}}{\nu_{\rm e}} & = & \left\{
\begin{array}{lll}
\frac{7}{4} & & \varepsilon^{\rm (I)}\approx 0.6275 , \cr
\frac{5}{3} & & \varepsilon\to\infty ,
\end{array} \right. \\
\delta_{\rm e} & = & \left\{
\begin{array}{lll}
15 & & \varepsilon^{\rm (I)}\approx 0.6275 , \cr
11 & & \varepsilon\to\infty ,
\end{array} \right. \\
\eta_{\rm e} & = & \left\{
\begin{array}{lll}
\frac{1}{4} & & \varepsilon^{\rm (I)}\approx 0.6275 , \cr
\frac{1}{3} & & \varepsilon\to\infty .
\end{array} \right.
\end{eqnarray} 
The fact that $\gamma_{\rm e}/\nu_{\rm e}$ and $\delta_{\rm e},\eta_{\rm e}$ 
are different at the two exactly solvable cases supports the full 
nonuniversality of the symmetric 16-vertex model on the square lattice.

\section{Conclusion} \label{Sec5}
The system under consideration was the symmetric two-state 
16-vertex model on the square lattice. 
Its vertex weights, which are invariant under any permutation of adjacent edge 
states, are considered to be symmetric with respect to the flip of all 
adjacent edge states $(+)\leftrightarrow (-)$ 
(see Fig.~\ref{fig:vertexweights}).
Such vertex weights automatically lie on the self-dual manifold of the gauge 
transformation (\ref{selfdual}), i.e., the subspace of the parameter
space which contains all the critical points.
The order parameter is the mean polarization $P$, see 
Eq. (\ref{polarization}).
The parametrization of vertex weights (\ref{spinflip}) contains two positive 
parameters, the temperature $T$ and the energy $\varepsilon$.
The two exactly solvable cases, namely the Ising model and 
the specific version of Baxter's 8-vertex model correspond to 
$\varepsilon^{\rm(I)}\approx 0.6275$ and $\varepsilon\to\infty$, respectively.
To study the critical properties of the model, we have applied 
the very accurate CTMRG method.
The dependence of the critical temperature $T_c$ on $\varepsilon$
is pictured in Fig.~\ref{fig:criticaltemperature}.
The fit of the plot in the region of small $\varepsilon$ (see the inset) 
indicates the linear dependence with $T_c$ going to $0$ as $\varepsilon\to 0$.
The plot of the critical exponent $\beta_{\rm e}$ versus $\varepsilon$, 
calculated with optimal parameters of the temperature step $\Delta T = 10^{-4}$
and dimension of the reduced space $m=200$, is represented by squares 
in Fig.~\ref{fig:betae}.
The critical exponent $\delta_{\rm e}(\varepsilon)$ is calculated with optimal 
parameters of the electric field $E=2.5\times 10^{-5}$ and $m=100$, see 
triangles in Fig.~\ref{fig:deltae}. 
The plots of the exponent $\gamma_{\rm e}$ versus $\varepsilon$ are evaluated 
``from first principles'' (triangles) and by using the scaling relation 
(\ref{scalinga}) (circles) in Fig.~\ref{fig:gammae}.
The dependence of the critical exponents $\nu_{\rm e}$ and $\eta_{\rm e}$
on $\varepsilon$, evaluated by (\ref{scalingb}) and (\ref{scalingc}), 
are presented in Fig.~\ref{fig:nue}.
All the critical exponents interpolate correctly between their 
known values at the two solvable cases $\varepsilon^{\rm(I)}\approx 0.6275$ and 
$\varepsilon\to\infty$.
The continuous variation of the critical exponents with the model 
parameter $\varepsilon$ is such that the weak universality hypothesis 
is violated.

\begin{acknowledgments}
The support received from the project EXSES APVV-16-0186 and VEGA Grants 
Nos. 2/0003/18  and 2/0123/19 is acknowledged. 
\end{acknowledgments}

\end{document}